УДК 338.242.42

# ПРИОРИТЕТЫ ИНСТИТУЦИОНАЛЬНОГО РЕГУЛИРОВАНИЯ ЭЛЕКТРОННОЙ КОММЕРЦИИ: РОССИЯ И МИРОВЫЕ ТЕНДЕНЦИИ
*Калужский М.Л.*

***Аннотация:*** Электронная коммерция постепенно трансформируется из разновидности торговой деятельности в самостоятельную отрасль глобальной сетевой экономики, которую невозможно игнорировать. Российская Федерация лидирует в СНГ по развитию электронной коммерции, но отстает от мировых лидеров в её институционализации. В статье анализируются проблемы государственного регулирования электронной коммерции в России и предлагаются пути их решения.

***Ключевые слова:*** электронная коммерция, сетевая экономика, государственное регулирование, электронная самозанятость, институциональная среда, экономический кризис.

# PRIORITIES OF INSTITUTIONAL REGULATION OF E-COMMERCE: RUSSIA AND WORLD TRENDS
*Kaluzhsky M.L.*

***Abstract:*** E-commerce is gradually transformed from a version of trading activity to independent branch of global network economy which cannot be ignored. The Russian Federation is in the lead in the CIS on development of e-commerce, but lags behind world leaders in institutionalization of e-commerce. Problems of state regulation of e-commerce in Russia are analyzed in article, ways of their decision are offered.

Регулирующая роль государства является частью институционализации электронной коммерции и одним из факторов формирования маркетинговой среды глобальной сетевой экономики. Задача государства заключается в том, чтобы определить законодательные нормы и правила электронной коммерции, а также вывести её из тени через ускоренное создание сетевой инфраструктуры. Если инфраструктура электронной коммерции не будет создана в России в ближайшие годы, то на российском потребительском рынке будет доминировать зарубежные сети товародвижения.

**Структурные изменения в экономике**. Развитие электронной коммерции вызывает неизбежные структурные изменения в экономике. В виртуальной среде происходит формирование транснациональных кластеров, концентрирующих «*интеллектуальные и инновационные отрасли и развивающие связи с другими производителями и клиентами*» [28, С. 172]. Эти кластеры существуют часто вне юрисдикции отдельных государств, обладая несравнимо большей конкурентоспособностью в сравнении с традиционными экономическими субъектами.

Причина кроется в большем доступе к рынкам и меньших трансакционных издержках. Например, продавец из США виртуально торгует на интернет-аукционе «eBay» с прямой отгрузкой товаров из Гонконга, где находится китайский поставщик, получающий товар с японского завода в Шэньчжэне. Или, например, российская фирма заказывает изготовление партии товара в Китае для последующей прямой поставки своим партнёрам на постсоветском пространстве. Мировые экономические лидеры при этом остаются прежними, так как «*основная часть мировой экономической активности и международной торговли сосредоточена в трёх крупных географических регионах: Северной Америке, Тихоокеанской области/Китае (включая Японию) и Западной Европе.*

*На эти три региона приходится 80% мирового экономического продукта и 75% мирового экспорта»* [26, С. 524].

Изменяется причина их конкурентоспособности. Если в традиционной экономике она была связана с производственным потенциалом и технологическими разработками, то в сетевой экономике конкурентоспособность определяется развитием инфраструктуры электронной коммерции. Бизнес уходит в Интернет и его контролирует тот, кто контролирует сетевую инфраструктуру. Уже не столь важно, где расположены производственные мощности, или какие технологии используются при производстве товаров массового спроса. Всё это можно купить, скопировать или создать самостоятельно. Однако без доступа к глобальному рынку, который представляет электронная коммерция, эти традиционные экономические факторы конкурентоспособности будут бесполезны. Поэтому традиционные отрасли производства трансформируются в глобальные отрасли, в которых *«конкурентные позиции компании на местном и национальном рынках определяются их глобальными позициями»* [20, С. 950].

Структурная трансформация экономических отношений ведёт к возникновению четырёх глобальных сдвигов в мировой экономике [2, С. 279]:

1. Рост глобального экономического значения бизнес-сетей, определяющих конкурентные преимущества на рынке.

2. Разделение информационных и товарных потоков, когда сделки глобализуются и могут одновременно совершаться по всему миру, независимо от местонахождения товара.

3. Расслоение мировой экономики, благодаря которому в ней формируются новые быстро растущие отрасли с повышенной доходностью.

4. Приоритетное развитие экономик с «возрастающими доходами» за счёт глобального перераспределения материальных и информационных потоков.

Это особенно актуально «*в условиях экономического спада*» и падения потребительского спроса, когда, как отмечает Ф. Котлер, основной задачей должна стать выработка «*мер по сокращению издержек*» [18, С. 177]. В результате массово деградирует и трансформируется традиционная оптово-розничная торговля, а трансакционные издержки, связанные с её деятельностью, перераспределяются между потребителями и логистическими провайдерами.

Ускоренное экономическое развитие происходит за счёт сокращения цепочек товародвижения и привлечения внешних ресурсов глобальных рынков. При этом рост электронной торговли во многом сопряжён со спадом в традиционной торговле. В мировой экономике наблюдается то, что С. Боулз назвал «институциональным вытеснением», которое «*происходит, когда присутствие одного института нарушает функционирование другого*» [9, С. 470]. Так, например, по состоянию только лишь на 2007 год «*начавшаяся электронная торговля в Японии уже оставила без работы более миллиона посредников разного уровня*» [21, С. 23].

В условиях виртуальной глобализации вопрос заключается лишь в том, в каких странах аккумулируются дивиденды от развития электронной коммерции, а в каких – убытки от сопутствующей деградации традиционной торговли. Не случайно многие западные экономисты прямо указывают на то, что электронная коммерция сегодня «*один из последних участков, где бизнес может получить преимущества, и место для будущей глобальной конкуренции*» [7, С. 542]. В числе приоритетов институционального развития розничной торговли они отмечают, что «*необходимо разработать информационные системы управления розничной торговлей, чтобы обеспечивать глобальные поставки, совершать платежи и выполнять все банковские требования*» [15, С. 64]. Именно здесь разворачивается глобальная конкуренция, которая не связана напрямую ни с производственными технологиями, ни с финансовым потенциалом.

Безусловным мировым лидером по темпам развития электронной коммерции является сегодня Китай. Первоначально одним из факторов феноменального роста китайской экономики стала «*открытость экономики (кайфан чжэцэ), основанная на экспортоори-* *ентированной модели развития, предполагающей за счет роста валютной выручки повышение техно- и наукоёмкости экономики, освоение новейших информационно-коммуникационных технологий, внедрение современных схем промышленной логистики»* [11, С. 11-12].

Однако одного лишь производственного потенциала явно недостаточно, если отсутствует распределительная инфраструктура. Поэтому сегодня «*стратегия Китая не столько использование пассивной оборонительной тактики торговой защиты, сколько активная наступательная тактика получения преимуществ за счет интеграционных союзов*» [11, С. 344]. Китай активно использует возможности электронной коммерции для трансграничного продвижения товаров на зарубежные рынки, как через прямые продажи, так и посредством развития дропшиппинга и других форм экономической кооперации.

Особенность китайского подхода к организации электронной коммерции заключается в активном регулирующем вмешательстве государства, когда государство ставит задачи перед бизнесом и развивает с его помощью приоритетные направления экономики. Политику китайского правительства в сфере электронной коммерции можно охарактеризовать тремя словами: направление, стимулирование и содействие. Государство в Китае определяет направление экономического развития, стимулирует экономическую активность и содействует субъектам экономики в достижении поставленных целей. Так, например, 08.01.2005 г. в Китае был опубликован документ под названием «Некоторые взгляды канцелярии Госсовета КНР на ускоряющееся развитие электронной коммерции». В нём содержался перечень институциональных мер, направленных на ускорение развития электронной коммерции [29, С. 94-95]:

1. Внесение изменений в политический курс и законодательную систему КНР, с целью усовершенствования норм права, финансовой системы и системы налогообложения, создание благоприятные условия для инвестиций.

2. Ускорение создания системы поддержки электронной коммерции в сфере кредитования, стандартов, платежей и перевозок.

3. Повышение доступности информации, а также пропаганда электронной коммерции среди крупных, средних, малых предприятий и потребителей.

4. Усовершенствование технической базы и системы обслуживания электронной коммерции.

5. Пропаганда необходимости подготовки кадров в сфере электронной коммерции.

6. Расширение международного сотрудничества, связанного с электронной коммерцией.

7. Содействие участию электронного бизнеса в международных торгово-промышленных выставках.

При этом речь идёт не только продвижении китайских товаров, но и о решении глобальных задач. В 2010 г. Китай совместно с Брунеем, Индонезией, Малайзией, Филиппинами, Сингапуром и Таиландом создали зону свободной торговли CAFTA (*China and ASEAN free trade area*), на базе стран – членов АСЕАН. В условиях недостаточного развития традиционных каналов товародвижения именно электронная коммерция является одним из тех механизмов, которые обеспечивают прирост взаимной торговли CAFTA примерно на 20% ежегодно [11, С. 345].

Россия также создаёт Таможенный союз на территории СНГ, активно вовлекая в него государства содружества. Проблема заключается в том, что Россия не располагает экономическими возможностями Китая и может предложить партнёрам в основном энергоносители и сырьё низкой степени переработки. Поэтому для достижения поставленной цели ей требуется «*принципиально иной сценарий перехода к рынку – не в результате плавного развития, а в форме резкого скачка*» [14, С. 118]. Для того чтобы реализовать такой сценарий, необходимо определить стратегические направления экономического развития, цели и пути их достижения. В этой сфере требуется целенаправленная политика государства.

Вместе с тем, в Концепции долгосрочного социально-экономического развития РФ на период до 2020 г. лишь вскользь упоминается о необходимости создания «*условий для развития компаний, работающих в области электронной торговли*» в качестве мер по обеспечению конкурентоспособности и технологического развития информационно-коммуникационных технологий [16]. Одновременно в качестве одной из приоритетных областей определено сотрудничество с Японией и Республикой Корея как источника для получения технологий. Для этого предполагается создание в партнерстве с третьими странами интегрированной транспортной и логистической инфраструктуры в Северо-Восточной Азии, развитие сотрудничества в области транспорта, прежде всего в сфере транзитных перевозок.

Транзитные перевозки – это важно, но не они влияют на рост промышленного производства, особенно в сфере потребительских товаров. Не эти страны являются сегодня лидерами в глобальной сетевой экономике, а Китай (особенно Гонконг) и США [6, С. 143-145]. Кроме того, продукция российских товаропроизводителей по понятным причинам не востребована пока ни в Южной Корее, ни в Японии, что автоматически превращает Россию в страну, зарабатывающую на транзите.

Действительно, «*для современных постиндустриальных экономик, ... прямые и широкие по охвату инструменты государственной поддержки уступают место узким, в том числе направленным не непосредственно на промышленность, а на обеспечивающую инфраструктуру*» [13, С. 137]. Однако вопрос заключается в том, какое промышленное производство будет обеспечивать будущая транспортно-логистическая инфраструктура. Только в Китае существует т.н. «производство на заказ». Южная Корея ориентирована на корпоративное производство, а японские корпорации сами перевели свои производственные мощности и заказы в Китай [11, С. 337].

Переориентация Китая на стимулирование внутреннего спроса и повышение курса юаня привело к тому, что заработная плата на промышленных предприятиях в Китае уже приближается к российским показателям. Основное конкурентное преимущество китайской экономики, заключающееся в дешевой рабочей силе, постепенно утрачивается. Поэтому Китай делает ставку на ускоренное развитие инновационных технологий, так как в традиционных технологиях он вскоре утратит ценовое преимущество.

Россия, в свою очередь, может сделать ставку на инновационную инфраструктуру распределения товаров на постсоветском пространстве. Тогда удастся удержать эти рынки хотя бы до того момента, когда промышленное производство потребительских товаров станет рентабельным и на рынке появятся отечественные товары массового спроса. Для обеспечения интеграционных процессов на постсоветском пространстве Россия могла бы предложить потребительскому рынку пусть не товары массового спроса, но инфраструктуру продвижения этих товаров. Товары могут быть китайскими или изготавливаться в Китае по заказу российских производителей. Они могут быть японскими или южнокорейскими. Проблема лишь в том, что трансграничная распределительная логистика уровней 4PL и 5PL существует пока в Китае и США.

Вместо ставки на привлечение в развитие Дальнего Востока инвестиций из США, Японии и Южной Кореи с целью ограничения экономической экспансии Китая, Россия могла бы сделать ставку на создание распределительной инфраструктуры электронной коммерции. Через эту инфраструктуру можно направить товарные потоки из Китая и той же Японии, которые сегодня идут помимо Российской Федерации. Эта инфраструктура должна быть способна одновременно проводить товары из-за рубежа и из России на рынки СНГ, на практике формируя единого экономического пространства. В распределительной логистике у России пока ещё сохраняется важное преимущество перед зарубежными конкурентами – единое русскоязычное пространство как основа для экономической интеграции. Если сделать ставку на развитие распределительной инфраструктуры электронной коммерции, а не на транзитные перевозки, то это позволит создать базу для интеграционных экономических процессов в рамках ЕврАзЕС и Таможенного союза.

Особое внимание следует обратить на опыт Китая в строительстве производственно-сбытовой инфраструктуры сетевой экономики, которая не просто является одной из ведущих в мире. Гораздо важнее то, что эта инфраструктура носит глобальный характер и развивается вне зависимости от содействия или противодействия органов госуправления в России. Уже сегодня китайские производители определяют экономическую конъюнктуру на потребительских рынках и в России, и на постсоветском пространстве. Речь идёт о том, использует российская экономика выгоды и преимущества от координации и интеграции с китайской экономикой или и далее будет делать ставку на политику институционального самоизоляционизма. С учётом вступления России в ВТО, развития электронной коммерции и глобализации мировых рынков игнорирование потенциала российско-китайского экономического сотрудничества негативно отражается только на российской экономике.

Задача состоит в том, чтобы интегрироваться в распределительную инфраструктуру сетевой экономики, которая уже активно формируется на постсоветском пространстве. Для России *«решение могло бы находиться в следующей плоскости: создание необходимых институциональных рамочных условий, без которых никакие инновационные возможности не могут быть реализованы»* [5, С. 37]. Вопрос стоит ребром: либо Россия в полной мере воспользуется возможностями и преимуществами сетевой экономики для решения задач модернизации экономики, либо упустит представившийся ей исторический шанс.

Следует отметить, что на уровне отдельных министерств и ведомств существует понимание важности обозначенных проблем. Так, например, в Прогнозе долгосрочного социально-экономического развития до 2030 г., составленном Минэкономразвития РФ, прямо говорится том, что *«рост производительности труда в торговле и некоторых других отраслях будет обеспечен переходом на новые формы производства (интернет-торговля, другие виды электронных услуг)»* [23, С. 71]. Однако на уровне Правительства РФ вопросам развития электронной коммерции, а также сопутствующей ей платёжной и распределительной инфраструктуры, пока не уделяется должного внимания.

**Нормативно-правовое регулирование**. Необходимость нормативно-правового регулирования электронной коммерции, прежде всего, связано с изменением институционального содержания экономических отношений. Изменения экзогенно обуславливают *«появление внутри общества большого числа индивидов, которые действуют с нарушением соглашения, что в результате и приводит к смене соглашения»* [9, С. 353]. Экономические отношения в электронной коммерции уже достаточно далеко вышли за институциональные рамки правового регулирования. Пришла пора трансформировать их нормативно-правовую базу в соответствии с изменившимися реалиями сетевого рынка.

Проблема состоит в том, что *«в современном налоговом законодательстве не существует эффективных методов налогового контроля, которые могли бы использоваться в целях выявления субъектов электронной коммерции, уклоняющихся от постановки на налоговый учёт, либо занижающих величину фактически полученных доходов»* [17, С. 5]. Это относится не только к России, но и практически ко всем без исключения другим странам мира. Актуальность проблемы оппортунистического поведения налогоплательщиков сравнительно невелика в секторе B2B. Однако в секторе B2C *«процент сокрытия фактических сделок может достигать 50-80%»* [17, С. 13]. При этом сектор C2C вообще находится вне какого-либо налогообложения, тогда как львиная доля продаж в электронной коммерции приходится именно на него. Достаточно сказать, что 90-95% дропшипинговых сделок происходит сегодня в секторе C2C, где оплата поступает на обезличенные счета, а товар отгружается покупателю третьим лицом в качестве подарка.

Не случайно американский экономист В. Танзи выделяет электронную коммерцию, электронные платежи и покупки за границей в качестве трёх из восьми «фискальных термитов», разъедающих основания налоговых систем [3, С. 4-15]. Можно выделить целый ряд институциональных особенностей электронной коммерции, затрудняющих её налогообложение и нормативно-правовое регулирование:

1) экстерриториальность электронных сделок в виртуальном пространстве, где отсутствуют географические границы;

2) анонимность продавцов и нематериальность виртуальных представительств, не позволяющая идентифицировать их местонахождение, так как сайты могут быть зарегистрированы в любой стране мира;

3) анонимность электронных сделок, так как платёжные системы проводят платежи за товары как частные денежные переводы (банки), либо вообще находятся вне правового поля (платёжные провайдеры);

4) анонимность клиентов, не позволяющая продавцу точно определить правовой статус клиента (физическое или юридическое лицо, цель покупки и т.д.).

В условиях глобальной конкуренции электронная коммерция позволяет *«неограниченно наращивать продажи в стране, не имея в ней никакого физического присутствия. Это приводит к росту налоговой базы в странах-экспортерах услуг и технологий, таких как США, за счёт сокращения этой базы в странах-импортерах»* [17, С. 25].

Вместе с тем, это не означает, что электронная коммерция имеет лишь негативный оттенок, так как речь тут идёт не об экспорте товаров, а об экспорте услуг распределительной логистики и технологий ведения трансграничного бизнеса. Их особенность заключается в относительной дешевизне внедрения и высокой экономической отдаче. Поэтому стратегическая задача государственного регулирования электронной коммерции в России заключается в том, чтобы трансформировать её распределительную инфраструктуру из импортно-ориентированной в экпортно-ориентированную. Причём, в условиях засилья зарубежных товаров на местных рынках, речь может идти и о приоритетном вытеснении зарубежных систем распределения.

Необходимый зарубежный опыт в этой сфере уже наработан. Электронный бизнес в виртуальной среде транснационален и трансграничен. Он де-факто находится не там, где располагаются покупатели, посредники и продавцы, а там, где располагается торговая инфраструктура электронной коммерции – т.е. торговая площадка. Участники виртуального бизнеса легко могут мигрировать из одной юрисдикции в другую, тогда как электронная торговая площадка привязана к месту регистрации. Поэтому, например, Китай в развитии электронной коммерции сделал ставку на национальную платёжную систему «UnionPay», национальных распределительных и торговых провайдеров.

В целом, если ориентироваться на зарубежный опыт институционализации электронной коммерции, можно выделить три основных модели её институционального регулирования: европейскую, американскую и китайскую.

***Европейская модель*** предполагает тотальную регламентацию и регистрацию субъектов электронной коммерции и совершаемых ими сделок. Примером такой деятельности может служить введение ведущими европейскими странами (Германия, Франция и Швейцария) общедоступного реестра добросовестных продавцов, зарегистрированных в налоговых органах. На сайтах с товарными предложениями продавцы обязаны указывать идентификационный номер своей государственной регистрации.

При этом нельзя сказать, что европейская модель оправдывает себя, так как, например, в сфере C2C европейская торговая инфраструктура не является мировым лидером в электронной коммерции из-за институциональных ограничений. Выигрывают от неё в основном крупные традиционные товаропроизводители и розничные торговые сети за счёт ограничения внешней конкуренции.

***Американская модель*** предполагает практически полный отказ государства от вмешательства в сферу электронной коммерции *«с целью максимизации выгод от использования экономического потенциала сети для национальных экономик»* [17, С. 19]. В основе этой модели лежит приятый в 1998 г. на три года билль «О налоговой свободе в интернете». Этот нормативный акт призван *«Определить национальную политику против государственного и местного вмешательства в межгосударственную торговлю на Интернет-сервисах или онлайн-сервисах, и ограничить юрисдикцию конгресса по межгосудар-* *ственной торговле, установив мораторий на наложение требований, которые вмешались бы в свободный поток торговли через Интернет*» [1, С. 1]. С тех пор установленный мораторий регулярно продляется, вплоть до настоящего времени.

Сущность американской модели регулирования электронной коммерции заключается в том, чтобы создать институциональные условия для концентрации и приоритетного развития её инфраструктуры на территории США. Не случайно крупнейшие глобальные торговые площадки (Amazon, eBay) и платёжные провайдеры (PayPal) с многомиллиардными оборотами находятся именно там.

***Китайская модель*** предполагает приоритет институционального развития электронной коммерции как инструмента продвижения китайских товаров на внешние рынки и развития распределительной инфраструктуры в самом Китае. Электронная коммерция рассматривается в Китае не столько как источник налоговых поступлений, сколько как стратегически важный механизм стимулирования промышленного производства. Именно поэтому «*в Китае большинство логистических видов деятельности управляется органами власти или строго контролируются ими*» [26, С. 526-527].

В отличие от других стран, в Китае для развития электронной коммерции государством целенаправленно создаются благоприятные институциональные условия: от отсутствия налогов до современной системы товародвижения и льготного таможенного режима. Налоги китайская экономика получает не с электронной коммерции (B2C и C2C), а с промышленного производства. Вместе с тем, нельзя сказать, что в Китае ничего не предпринимается для налогообложения предпринимателей в электронной коммерции. Планируемая налоговая реформа сделает личный подоходный налог «*одним из основных видов налогов*», который «*будет играть более важную роль, чем налог на доходы предприятий*» [11, С. 210].

Таким образом, в сфере институционального регулирования электронной коммерции можно обнаружить весь спектр подходов: от жёсткой регламентации до попустительства и сознательного стимулирования. Лишь в одном все подходы к институционализации электронной коммерции едины – в понимании того, что она представляет собой глобальное явление, которое должно регулироваться глобальными нормами. В институциональном смысле электронная коммерция пока не обрела ещё свою правовую категорию, и её нормативно-правовой статус окончательно не определён. Это проблема не только российской системы права, но международного права в целом [10, С. 2-5].

Вместе с тем, институциональные аспекты регулирования электронной коммерции не сводятся исключительно к налогообложению. Электронная коммерция находится сегодня в авангарде цивилизационного развития, обеспечивая беспрецедентный рост предпринимательской активности по всему миру. Тут сложно не согласиться с Ф. Котлером, который пишет: «*Основной, связанный с законодательством о предпринимательской деятельности вопрос заключается в следующем: когда затраты на регулирование начинают превышать выгоды? Каждый новый закон, бесспорно, может быть юридически оправдан, но одновременно с этим существует вероятность того, что его принятие приведет к ослаблению предпринимательской инициативы и замедлению экономического роста*» [19, С. 149].

Электронная коммерция обладает достаточно высокой эластичностью. По оценке И.А. Стрелец, применение к электронной коммерции существующих в России налоговых ставок способно привести к сокращению числа покупателей в ней на 20-25%, а объёмов продаж – на 25-30%, тогда как «*влияние сетевых сделок на поступления в бюджет может оказаться несущественным*» [27, С. 28]. Это, прежде всего, связано с тем, что сегодня в России «*удельный вес неуплаченных субъектами электронной коммерции налогов составляет величину в пределах 0,2% от совокупных налоговых поступлений*», включая составляющий львиную долю продаж сектор B2B [17, С. 13-14]. С другой стороны, как отмечается в Плане деятельности Министерства связи и массовых коммуникаций РФ на период 2013-2018 гг., «*Благодаря интернету развивается малый бизнес, электронная* *коммерция, растет производительность труда и эффективность бизнес-процессов предприятий ... . Каждые 10% проникновения быстрого и качественного интернета могут дать экономике рост ВВП на 1,4% в год»* [22, С. 15].

Решение задачи налогообложения субъектов электронной коммерции в России усложняется неспособностью *«налоговых органов контролировать соблюдение законодательства о налогах и сборах, а также правильность исчисления и уплаты налогов»* [17, С. 92]. При этом действующее законодательство Российской Федерации регулирует электронную коммерцию крайне противоречиво и далеко не всегда адекватно:

1. Закон Российской Федерации от 28 декабря 2009 г. № 381-ФЗ «Об основах государственного регулирования торговой деятельности в Российской Федерации» вообще не содержит ни одного упоминания об электронной коммерции, исключая её из сферы розничной торговли.

2. Согласно п. 2 ст. 309, пп. 3 и 8 ст. 306 и ст. 146 Налогового Кодекса РФ, доходы, полученные иностранной организацией от продажи товаров, имущественных прав, работ и услуг на территории Российской Федерации, не приводящие к образованию постоянного представительства, обложению налогами (НДС, налогом на прибыль и т.д.) у источника выплаты не подлежат.

3. Согласно инструктивным письмам Министерства финансов РФ № 03-11-02/86, № 04-05-12/20 и № 04-05-11/50 торговые предприятия при переносе продаж в Интернет не вправе использовать режим единого налога на вмененный доход и должны вести раздельный учет. Это означает, что издержки электронной коммерции у продавцов гораздо выше, чем при осуществлении стационарных продаж, не говоря уже о том, что зарубежные продавцы вообще освобождены от всех налогов.

Действующее законодательство рассматривает электронную коммерцию в качестве обычной предпринимательской деятельности, применяя к ней соответствующие нормы, не имея физической возможности обеспечить их соблюдение на просторах Интернета. В результате *«российская практика налогообложения ... не учитывает уникальные особенности электронной коммерции в экономике РФ. Большинство представителей электронного бизнеса активно используют пробелы гражданского и налогового законодательства и с лёгкостью обходят основные положения НК РФ»* [17, С. 34].

Теневое развитие электронной коммерции происходит до определённого предела роста, за которым её субъекты всё-таки вынуждены легализовывать свою деятельность. Только делают они это уже за рамками юрисдикции Российской Федерации, чаще всего в оффшорных зонах. Так институциональное несовершенство законодательства искусственно ограничивает развитие национальной инфраструктуры электронной коммерции, создавая неоправданные преференции для её зарубежных участников.

Для исправления сложившегося положения требуется целенаправленная государственная политика по развитию электронной коммерции, которая должная стать одним из стратегических направлений экономического развития страны в ближайшее десятилетие. Этого нельзя добиться одними запретительными мерами или игнорированием происходящих на виртуальном рынке процессов. Государство должно принять институциональные меры по легализации электронной коммерции и превращению её в один из локомотивов модернизации экономики. Как совершенно справедливо отмечает С. Боулз: *«Чтобы принуждать и при этом предотвращать бегство от принуждения, государство должно быть универсальным и непревзойденным в некоторых отношениях»* [9, С. 464]. В качестве возможных вариантов решения проблемы можно предложить следующие меры:

1. Рассмотреть вопрос о принятии закона вводящего мораторий на налогообложение субъектов электронной коммерции по аналогии с биллем «О налоговой свободе в интернете» в США. Одновременно можно ввести уведомительную государственную регистрацию этих субъектов. Это позволит легализовать электронную коммерцию, создав условия для её ускоренного развития.

2. Обеспечить правовое закрепление институциональной нормы электронной коммерции, впервые сформулированной Бюро цензов Министерства торговли США, согласно которой: «*Соглашение в электронной форме, а не платеж, является ключевым конституирующим признаком сделки в сфере электронной торговли*» [17, С. 9]. Это сделает государство гарантом и полноценным участником институциональных отношений в электронной коммерции.

3. При переходе к налогообложению индивидуальных субъектов электронной коммерции использовать минимизированный Единый налог на вмененный доход, либо, по аналогии с Китаем, ограничиться взиманием подоходного налога. Это позволит активизировать социальную роль электронной коммерции, превратив её в одну из форм дистанционной занятости, что особенно актуально в условиях развивающегося экономического кризиса.

4. Внедрить меры по защите прав покупателей приобретающих товары у зарегистрированных субъектов электронной коммерции по аналогии с европейскими странами. Это послужит важным стимулом для добровольной регистрации участников виртуального рынка.

5. При таможенном оформлении грузов перейти от декларирования партий товара к декларированию видов товара. Например, в Германии это делается в форме выдачи лицензий на импорт продукции, которую могут получить как юридические, так и физические лица. После получения лицензии разрешительное декларирование товара заменяется уведомительным декларированием поставок, где за достоверность данных ответственность несёт грузоотправитель.

Цель такого регулирования заключается в приоритетном развитии электронной коммерции, а не в её ограничении. Электронная коммерция не способна принести большие бюджетные доходы. Её экономическая функция состоит в создании условий для продвижения промышленной продукции на внутренний и зарубежные рынки. Задача заключается в заимствовании лучшего из мирового опыта: в Европе – правовой защиты покупателей на основе электронной подписи и регистрации продавцов, в США – законодательного моратория на налогообложение электронных сделок, в Китае – мер по формированию распределительной инфраструктуры уровней 4PL и 5PL, а также адресной поддержки провайдеров торговых и распределительных сетей.

**Социально-экономическая политика**. Электронная коммерция имеет не только важное экономическое, но и не менее важное социальное измерение. Виртуализация рынка труда провоцирует развитие удалённой занятости, постепенно изменяя его структуру. Это ведёт к повышению виртуальной мобильности населения и снижению территориальных диспропорций, что особенно актуально в условиях глобального экономического кризиса и продолжающегося падения промышленного производства.

Одновременно происходят весьма существенные изменения в социальных отношениях. Профессиональные интернет-сервисы и торговые площадки становятся центрами межличностной социальной интеграции по профессиональному признаку. Они не просто предоставляют пользователям возможность интерактивного общения и социальный статус в сети, но и обеспечивают средствами к существованию, обучают, формируют виртуальную среду обитания.

Пока это делается по-партизански. Государство практически не участвует в формировании виртуальной социальной среды, институциональные нормы и правила которой определяются провайдерами торговых, платёжных или коммуникативных услуг. При этом уникальная особенность виртуальной среды заключается не только в автономности, но и в высшей степени интеграции её коммуникативных и коммерческих функций.

Провайдеры услуг, связанных с электронной коммерцией, самостоятельно устанавливают правила поведения участников, за нарушение которых приостанавливают или аннулируют аккаунты субъектов электронной коммерции и покупателей. Вместе с тем, они не способны самостоятельно предпринять юридические санкции в отношении нарушите-лей и здесь располагается большое поле деятельности для государственного регулирования. Если государство возьмёт на себя функцию защиты прав участников электронных сделок так же, как оно уже это делает в сфере банковского кредитования, то задача его легитимизации будет решена.

Государство не сможет институционализировать электронную коммерцию, пока не станет гарантом связанных с ней отношений. Применительно к социально-экономическому регулированию речь идёт о легализации электронной самозанятости, которая *«практически полностью находится вне правового поля и относится к неформальной экономике. Помимо низкого уровня правовой культуры, в России это объясняется … высокой технологичностью самих рынков, многие аспекты которых в настоящее время не регулируются законодательно»* [25, С. 112].

Взаимосвязь электронной коммерции и электронной самозанятости предполагает принципиально новые возможности решения наиболее значимых социальных задач. В первую очередь это касается вопросов занятости, образования, развития депрессивных территорий и поддержки предпринимательства. Не случайно американский билль «О налоговой свободе в интернете» гласит: «(9) *Электронный рынок услуг, продуктов, и идей, доступных через Интернет, может быть особенно выгодным для пенсионеров, инвалидов, сельских жителей и предприятий малого бизнеса, а также предполагает новые возможности и льготы для образовательных учреждений и благотворительных организаций»* [1, С. 4].

Самозанятость в электронной коммерции позволяет решить целый ряд социальных проблем, присущих российской экономике:

1. Уменьшить негативный эффект от старения населения при сокращении экономически активной его части за счёт расширения трудовых ресурсов и создания виртуальных рабочих мест.

2. Мобилизовать предпринимательский потенциал безработных, пенсионеров, инвалидов и других групп населения для ведения активной экономической деятельности.

3. Снизить социальную напряжённость в обществе, связанную с обусловленным экономическим кризисом спадом предпринимательской активности и ростом безработицы.

4. Обойтись минимальными бюджетными расходами, получив долгосрочный общеэкономический эффект, мультиплицируемый за счёт привлекаемых с помощью электронной коммерции ресурсов.

При этом крайне важно, что *«электронная самозанятость стоит особняком в общем ряду нестандартных форм занятости, большинство которых в российской экономике относятся к вторичному рынку труда с низкой заработной платой и примитивными технологиями»* [25, С. 112]. В отличие от них она способна обеспечить достаточно высокий индивидуальный доход. Например, используемые в период острых экономических кризисов общественные работы неспособны даже близко обеспечить аналогичный социально-экономический эффект.

Весь исторический опыт свидетельствует о том, что *«на понижательной фазе <экономического цикла>, при которой падает спрос на рабочую силу и увеличивается число циклических безработных, как правило, возникает настоятельная необходимость в увеличении государственных расходов, связанных с программами на рынке труда»* [24, С. 113]. Поэтому особенно остро сегодня стоит вопрос о выборе приоритетных направлений целенаправленной социально-экономической политики государства, большую роль в которой может и должна сыграть электронная самозанятость на сетевом рынке.

Рынок труда электронной самозанятости в России обладает значительным потенциалом, который может в перспективе стать основой для формирования новых конкурентных преимуществ российской экономики. Так, перепись фрилансеров, проведённая в 2009-2011 гг. НИУ ВШЭ показала, что более 2/3 опрошенных (69%) из 34 стран мира русскоязычных фрилансеров находятся сегодня в России. При этом «*на россий-* *ских заказчиков работают 72% фрилансеров из других стран ... и 86% респондентов в целом»* [25, С. 99].

Результаты проведённого НИУ ВШЭ исследования свидетельствует о лидирующей роли фрилансеров из России на постсоветском пространстве, что уже делает Россию центром интеграционного притяжения. Россия лидирует в этом неформальном сегменте рынка труда, хотя и отстает от Европы и США. Однако без легализации электронной самозанятости и государственной поддержки направления лидирующее положение российских фрилансеров будет утрачено уже через несколько лет.

Данные НИУ ВШЭ свидетельствуют о том, что российская электронная самозанятость остро нуждается в формализации сложившихся институций, который невозможен без решения этой задачи. На постсоветском пространстве в России довольно давно уже наблюдается количественное опережение в показателях институционального развития электронной самозанятости, которое должно быть трансформировано в качественные институциональные изменения.

Общие тенденции развития рынка труда в России также свидетельствуют о необходимости выработки новых институциональных подходов к регулированию электронной самозанятости. Среди наиболее острых проблем социально-экономической политики здесь можно выделить следующее:

1. Отсутствие стратегий реструктуризации рынка труда в условиях глобального экономического кризиса.

2. Сконцентрированность на поддержании занятости в традиционной экономике в ущерб стимулированию электронной занятости в сетевой экономике.

3. Ориентация Трудового кодекса РФ на традиционную занятость индустриального типа, не подразумевающую возможности электронной занятости.

4. Фактическое игнорирование регулятором уже существующего в России виртуального рынка труда и электронной самозанятости населения.

Всё это негативно сказывается на гибкости рынка труда и ведёт не только к массовому нарушению норм трудового законодательства, но и к системному воспроизводству неформальной занятости. С другой стороны, игнорирование регулятором институциональных реалий сетевой экономики тормозит её институциональное развитие и провоцирует застойные явления и в традиционном секторе.

Существует прямая связь между развитием электронной самозанятости и развитием частного предпринимательства. Политика занятости не должна ограничиваться рекрутированием безработных граждан в низкооплачиваемые и низкоквалифицированные сферы труда. *«Цепь поставок – одна из последних крупных областей, где бизнес может получить преимущества»*, – совершенно справедливо отмечает Э.Дж. Бергер [7, С. 543]. Направляя предпринимательскую активность в сферу электронной коммерции, государство может решить целый ряд важных социально-экономических задач, не прибегая к отвлечению значительных финансовых ресурсов.

Не случайно именно электронная коммерция рассматривается сегодня во многих странах мира в качестве одного из важнейших механизмов обеспечения альтернативной занятости и развития частного предпринимательства. Для России это особенно актуально, поскольку согласно докладу Всемирного Банка «Doing Business» она занимала в 2013 г. лишь 112 место среди 185 стран в рейтинге благоприятности условий предпринимательской деятельности [4, С. 3]. Связанная с электронной коммерцией предпринимательская деятельность способна стать ещё одной немаловажной сферой обеспечения альтернативной занятости населения в России. Тем более что закон РФ от 19.04.1991 г. № 1032-1 «О занятости населения в Российской Федерации», в отличие от Трудового кодекса РФ, предоставляет для этого максимум возможностей [12]:

– Статья 5 определяет поддержку трудовой и предпринимательской инициативы граждан, осуществляемой в рамках законности, в качестве одного из направлений государственной политики содействия занятости населения;

– Статья 7 относит к полномочиям федеральных органов государственной власти координацию деятельности по созданию экономических условий занятости населения, развития предпринимательства и самозанятости;

– Статья 7.1 выделяет содействие самозанятости безработных граждан в качестве одной из государственных услуг, обязательных для оказания органами государственной власти субъектов Российской Федерации;

– Статья 10 подтверждает право граждан на профессиональную деятельность за пределами территории Российской Федерации, легализуя тем самым электронную занятость с регистрацией вне юрисдикции Российской Федерации.

Здесь особенно может быть полезен опыт зарубежных стран, осуществляющих государственные программы помощи безработным на основе стимулирования самозанятости и создания индивидуальных рабочих мест. Это позволяет им «*во время циклических депрессий смягчить социальные последствия возрастающего уровня безработицы путём реализации ряда учебных предпринимательских программ, направленных на помощь безработным в создании собственных индивидуальных предприятий без наёмных работников*» [8].

Отдельно следует упомянуть набирающую сегодня популярность в Европе датскую модель институционального регулирования рынка труда флексикьюрити («*flexicurity*»).[1] Суть её заключается в активной государственной политике занятости, подразумевающей гибкость и приспособляемость труда к меняющимся условиям рынка, а также социальную защиту, направленную на смягчение негативных последствий структурных изменений в экономике [24, С. 114]. С одной стороны, политика флексикьюрити направлена на обеспечение занятости и самостоятельно заработка экономически неактивной части населения. С другой стороны она способствует «*вымыванию неэффективных рабочих мест и перемещению освобождаемых работников в рентабельные и перспективные производства*» [24, С. 180]. Другими словами, речь идёт о переходе от субсидирования государством безработицы к содействию адаптации работников к меняющимся условиям рынка труда.

Самозанятость рассматривается в рамках флексикьюрити как важнейшая форма альтернативной занятости, не требующей привлечения стороннего работодателя. В европейском понимании она подразумевает осуществление собственного бизнес-проекта с работой только на себя без привлечения наёмных работников и независимо о расположения места работы. Применительно к электронной самозанятости такой подход формирует институциональную среду для укоренного развития виртуальных организаций, в которых отсутствует вертикальная инфраструктура управления. Не случайно именно Европа сегодня становится мировым лидером в этой области в отличии, например, от развития там электронной коммерции. Для России это является ещё одним свидетельством в пользу рамочного регулирования социально-экономического развития электронной коммерции.

Концепция долгосрочного социально-экономического развития Российской Федерации на период до 2020 г., утверждённая распоряжением Правительства РФ № 1662-р от 17.11.2008 г. также указывает на то, что «*инновационный тип экономического развития требует создания максимально благоприятных условий для предпринимательской инициативы, повышения конкурентоспособности и инвестиционной привлекательности российских частных компаний, расширения их способности к работе на открытых глобальных рынках в условиях жесткой конкуренции, поскольку именно частный бизнес является основной движущей силой экономического развития. Государство может создать необходимые условия и стимулы для развития бизнеса, но не должно подменять бизнес собственной активностью*» [16].

Формируемая государством институциональная маркетинговая среда электронной коммерции оказывает экзогенное влияние на реализацию экономического потенциала её субъектов. От их деятельности во многом зависит сегодня не только конкурентоспособ-ность российской экономики, но и судьба интеграционных процессов на постсоветском пространстве. Поэтому электронная коммерция, постепенно приобретая черты самостоятельной отрасли российской экономики, одновременно приобретает и стратегическое значение в качестве одного из источников её структурной модернизации.

---

[1] Сокращение от «**flexi***bility*» и «*se***curity**» (англ.).

*Библиографический список:*

1. Internet Tax Freedom Act of 1998, 105th Congress 2d Session, H.R. 3529 / U.S Government Printing Office (GPO).
2. *Koch R.* The Financial Times guide to strategy: how to create and pursue a winning strategy. 4th ed. London: Prentice Hall, 2011.
3. *Tanzi V.* Globalization and the Work of Fiscal Termites // IMF Working Paper 00/181. Washington: International Monetary Fund, 2000.
4. World Bank. Doing Business: 2013. Разумный подход к регулированию деятельности малых и средних предприятий. Washington (DC): World Bank Group, 2013.
5. *Айхелькраут С.* Институты, инновации и экономическая политика // Журнал институциональных исследований. 2009. № 1. Т. 1. С. 36 – 42.
6. *Бауэрсокс Д.Дж., Клосс Д.Дж.* Логистика: интегрированная цепь поставок. М.: Олимп-Бизнес, 2008.
7. *Бергер Э.Дж.* Е-коммерция и цепи поставок: ломка прежних границ / Управление цепями поставок. Под ред. Дж.Л. Гатторны. М.: Инфра-М, 2008. С. 540 – 555.
8. *Бондаренко В.А.* Развитие самозанятости в странах Европейского Союза и США. М.: МЦРП, 2009.
9. *Боулз С.* Микроэкономика. Поведение, институты и эволюция. М.: Дело, 2011.
10. *Васильева Н.М.* Электронная коммерция как правовая категория // Юрист. 2006. № 5. С. 2 – 7.
11. *Дин Жуджунь, Ковалев М.М., Новик В.В.* Феномен экономического развития Китая. – Мн.: БГУ, 2008.
12. Закон РФ от 19.04.1991 г. № 1032-1 «О занятости населения в Российской Федерации». М.: НПП «Гарант-сервис», 2013.
13. *Калинин А.М.* Построение сбалансированной промышленной политики: вопросы структурирования целей, задач, инструментов // Вопросы экономики. 2012. № 4. С. 132 – 146.
14. *Калинина А.Э.* Интернет-бизнес и электронная коммерция. Волгоград: ВолГУ, 2004.
15. *Кент Т., Омар О.* Розничная торговля. М.: Юнити-Дана, 2007.
16. Концепция долгосрочного социально-экономического развития Российской Федерации на период до 2020 г. / Утв. Распоряжением Правительства РФ № 1662-р от 17.11.2008 г. М.: НПП «Гарант-сервис», 2013.
17. *Корень А.В.* Налогообложение субъектов электронной коммерции: проблемы и перспективы. Владивосток: ВГУЭС, 2011.
18. *Котлер Ф.* 300 ключевых вопросов маркетинга: отвечает Филип Котлер. М.: Олимп-Бизнес, 2006.
19. *Котлер Ф.* Маркетинг менеджмент. Экспресс-курс. 2-е изд. СПб.: Питер, 2006.
20. *Котлер Ф., Армстронг Г.* Основы маркетинга. Профессиональное издание. М.: Вильямс, 2009.
21. *Крымский И.А., Павлов К.В.* Проблемы и перспективы развития электронной экономики в России. Мурманск: Кольский НЦ РАН, 2007.
22. План деятельности Министерства связи и массовых коммуникаций Российской Федерации на период 2013-2018 гг. М.: НПП «Гарант-сервис», 2013.
23. Прогноз долгосрочного социально-экономического развития Российской Федерации до 2030 года. М.: Минэкономразвития России, 2013.
24. Рынок труда: реакция на кризис (по материалам зарубежных стран) / Под ред. Ф.Э. Бурджалова, Е.Ш. Гонтмахера. М.: МЭМО РАН, 2011.

25. *Скребков Д., Шевчук А.* Электронная самозанятость в России // Вопросы экономики. 2011. № 10. С. 91 – 112.
26. *Сток Дж.Р., Ламберт Д.М.* Стратегическое управление логистикой. М.: Инфра-М, 2005.
27. *Стрелец И.А.* Модификация экономической политики государства в условиях новой экономики // Бизнес, менеджмент и право. 2005. № 3 (9). С. 22 – 30.
28. *Стрелец И.А.* Сетевая экономика. М.: Эксмо, 2006.
29. *Цзюнь Жун Х.* Развитие электронной коммерции в Китае / Электронная торговля в СНГ и восточноевропейских странах / Под. ред. Б.Н. Паньшина. Мн.: БГУ, 2006. С 92 – 95.